%% file: 00_main.tex
\documentclass[sigconf]{acmart}

\usepackage{listings}
\usepackage{array}
\usepackage{multirow}
\usepackage{framed}
\usepackage{blindtext}
\usepackage{url}
\usepackage{hyperref}
\usepackage{enumitem}
 {\endMakeFramed}
\definecolor{shadecolor}{gray}{0.92}

\copyrightyear{2026}
\acmYear{2026}
\setcopyright{cc}
\setcctype{by-nc-nd}
\acmConference[DIS '26]{Designing Interactive Systems Conference}{June 13--17, 2026}{Singapore, Singapore}
\acmBooktitle{Designing Interactive Systems Conference (DIS '26), June 13--17, 2026, Singapore, Singapore}
\acmDOI{10.1145/3800645.3813084}
\acmISBN{979-8-4007-2563-0/2026/06}

\usepackage{bbding}%
\AtBeginDocument{%
  \providecommand\BibTeX{{%
    \normalfont B\kern-0.5em{\scshape i\kern-0.25em b}\kern-0.8em\TeX}}}


\begin{document}

\title{Artistic Practice Opportunities in CST Evaluations: A Longitudinal Group Deployment of ArtKrit}

\newcommand{\commentText}[1]{#1}    
\newcommand{\TODO}[1]{\commentText{{\color{red}[\textbf{\textsc{TODO}}: \textit{#1}]}}}
\newcommand{\jiaju}[1]{\commentText{{\color{blue}[\textbf{\textsc{JM}}: \textit{#1}]}}}
\newcommand{\jingyi}[1]{\commentText{{\color{teal}[\textbf{\textsc{JL}}: \textit{#1}]}}}
\newcommand{\chauvu}[1]{\commentText{{\color{purple}[\textbf{\textsc{CV}}: \textit{#1}]}}}
\newcommand{\asya}[1]{\commentText{{\color{orange}[\textbf{\textsc{AV}}: \textit{#1}]}}}
\newcommand{\tao}[1]{\commentText{{\color{brown}[\textbf{\textsc{TL}}: \textit{#1}]}}}
\newcommand{\catherine}[1]{\commentText{{\color{magenta}[\textbf{\textsc{CL}}: \textit{#1}]}}}
\newcommand{\camready}[1]{\commentText{{{#1}}}}

\newcommand{\EDITED}[1]{{#1}}
\newcommand{\changes}[1]{{#1}} 

\renewcommand{\changes}[1]{{#1}}

\newcommand{\systemName}[1]{ArtKrit}


\author{Catherine Liu}
\authornote{Both authors contributed equally to this research.}
\affiliation{
  \institution{Claremont McKenna College}
  \city{Claremont}
    \state{California}
  \country{USA}
}

\author{Tao Long}
\authornotemark[1]
\affiliation{
  \institution{Columbia University}
  \city{New York}
    \state{New York}
  \country{USA}
}

\author{Asya Lyubavina}
\affiliation{
  \institution{Pomona College}
  \city{Claremont}
  \state{California}
  \country{USA}
}

\author{Chau Vu}
\affiliation{
  \institution{Pomona College}
  \city{Claremont}
  \state{California}
  \country{USA}
}

\author{Jiaju Ma}
\affiliation{
  \institution{Stanford University}
  \city{Stanford}
    \state{California}
  \country{USA}}

\author{Jingyi Li}
\affiliation{%
  \institution{Pomona College}
  \city{Claremont}
  \state{California}
  \country{USA}}

\begin{abstract}
Creativity support tools (CSTs) aim to elevate the quality of artists’ creative \changes{processes and artifacts}. 
Yet most current CST evaluations overlook temporal and social aspects of tool use. To address this gap, we present a longitudinal, group-based CST evaluation through a three-week deployment of ArtKrit, a computational drawing tool that supports disciplined drawing. Nine  digital artists, organized into three communities of practice, completed weekly ``master studies'' alongside a researcher-artist. Our results show users’ evolving relationships with ArtKrit over time---from early experimentation to selective incorporation or misuse---alongside changes in their ways of artistic seeing. These changes unfolded within artist support networks that fostered confidence and creative safety, and validated individual expression. 
Overall, our findings suggest that CST evaluations can---and should---be designed as opportunities for meaningful artistic engagement rather than purely extractive measurement exercises. \changes{We contribute this longitudinal, group-based approach as one CST evaluation method.}
\end{abstract}


\begin{teaserfigure}
  \includegraphics[width=\textwidth]{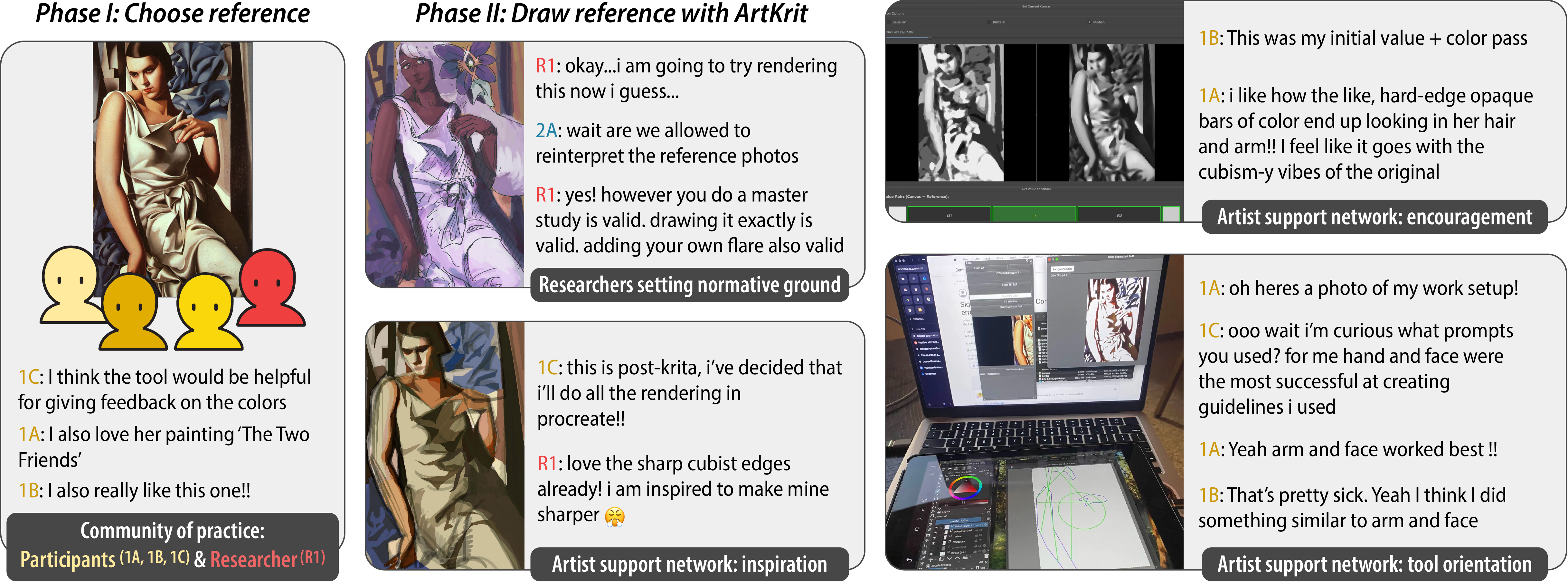}
  \caption{A sample of one group's week during our three-week evaluation. Nine participants were divided into three small communities of practice that included a member of the research team; each week, the group voted on and drew a ``master study''. Each group had their own Discord channel, visible to all participants. Participant comments are real Discord messages sent. Shown are works-in-progress shared from Week 1 of Group 1; Figure \ref{fig:group1final} shows final results. Our study structure encouraged building artist support networks \cite{chung-ast} and observing a mini creativity support ecosystem \cite{cse-shm}. \textit{Reference image: Portrait of Madame M. by Tamara De Lempicka.} 
  }
  \Description{A research study graphic showing two phases of online artistic collaboration among three participants (1A, 1B, 1C) and a researcher (R1). In Phase I, participants choose a painting reference. In Phase II, they share digital drawings and feedback in a Discord chat. Comments center around researchers setting normative ground, inspiration, encouragement, and tool orientation. The sample reference image used was "Portrait of Madame M." by Tamara de Lempicka.}
  \label{fig:teaser}
\end{teaserfigure}

\maketitle

\input{01_introduction}
\input{02_methods}

\input{03_findings}
\input{04_discussion}
\input{05_related_work}
\input{06_conclusion}

\begin{acks}
We owe a great debt to our study participants who made this research not only possible but also extremely fun. We also thank Jacob Ritchie, Shm Almeda, and Max Kreminski for their insightful conversations around this work, and CST evaluations broadly.
\end{acks}

\bibliographystyle{ACM-Reference-Format}
\bibliography{10_reference}

\end{document}

%% file: 01_introduction.tex
\section{Introduction}

Since establishing a Creativity Support Tools (CSTs) research agenda over 25 years ago \cite{shneiderman-UI-innovation}, human-computer interaction (HCI) researchers have grappled with how to most appropriately evaluate them. One challenge is that the goals of CSTs vary widely: as noted in a 2005 National Science Foundation (NSF) workshop organized by Shneiderman et al., ``the
problem of developing CSTs is one in which one must first decide in which intersection of the n-dimensional taxonomy one wishes to study and work'' \cite{hewett2005creativity}. The most common directions, as shown in a 2019 survey of CSTs \cite{frich_mapping_2019}, have been developing novel research prototypes for idea generation and artifact production. As such, researchers have used Likert scale surveys such as the NASA-TLX (Task Load Index) \cite{nasa-tlx} or the Creativity Support Index \cite{CSI} to operationalize their definitions of ``creativity''---often artifact- and process- centric---and justify that their tools help users achieve their ``creative'' goals. 

Yet, the authors of the 2005 CST workshop report that ``the first objective is to enhance the \textit{personal experience} of the person who wants to be creative,'' (emphasis added) while improving ``outcomes and artifacts'' and ``process'' are second and third objectives, respectively \cite{hewett2005creativity}. Most recently, Cox et al. highlight this oversight: beyond measures of productivity, HCI researchers should evaluate wholistic, \textit{user-centric} aspects, such as how CSTs promote self-reflection and emotional well-being \cite{beyondproductivity}.

The shift to evaluating how CSTs may support the experiences of artists, and not just the creation of artifacts, 
shares goals with Research through Design methodology \cite{gaver-rtd}, 
which generates knowledge through reflective engagement with design artifacts, particularly towards speculative, diverse, and sometimes convoluted encounters \cite{desjardins-rtd}. Similarly, researchers deploy technology probes not to measure tool effectiveness in controlled lab settings, but rather to ``find out about the unknown'' in real-world contexts \cite{tech-probe-family}. Commenting on the long-standing challenge of evaluating CSTs, Remy et al. advocate for longitudinal, in-situ studies, and coming together as a community to develop a toolbox of methods \cite{remy-evaluating-csts}.

This paper presents one such method for CST evaluation. Shaped by a belief that CST evaluations should surface generative insights about the creative practices and proclivities of the artists who use tools, we 
\changes{contribute a longitudinal, group-based evaluation methodology}, demonstrated through a three-week deployment
of ArtKrit \cite{artkrit} with nine experienced digital artists. ArtKrit is a Krita \cite{krita} plug-in that supports disciplined drawing, the process of studying and replicating reference images, also called ``master studies.'' ArtKrit offers computational guidance and feedback by decomposing reference images into composition, value, and color steps. 

\changes{In line with recent methodological developments like comparative structured observation \cite{CSO-mackay},} 
our goals were not to demonstrate ArtKrit's usability through statistical reproduction or to quantify the improvements in disciplined drawing  gained through using ArtKrit over a period of time. Instead, we used ArtKrit as a case study for a qualitative observation (supplemented with quantitative data logs) of \changes{the situations and ecosystems around disciplined drawing that ArtKrit exists in---}how computational scaffolds may alter digital artists' workflows, relationships, and ways of artistic seeing and interpretation over time.


Motivated by Chung et al.'s theory of artist support networks \cite{chung-ast}, and given the prevalent critique of the lack of group-based CST evaluations despite creativity being a social activity \cite{becker2023art}, we grouped our nine participants into three communities of practice \cite{cop-knowledge} on a shared Discord server. During each week of the three-week study, each community of practice collectively voted on a master study, which all members then drew individually. A member of the research team who was also an experienced digital artist joined each community of practice, drawing and posting on Discord. Our longitudinal, group-based study addresses a gap in prior CST evaluations that have largely relied on short-term, artifact-centric studies where participants do not know or interact with each other.

Our study aimed to answer three research questions:

\vspace{-0.5em}
\begin{enumerate}
    \item How do practitioners use ArtKrit in practice?
    \item How do practitioners' relationships with new creative software change over time?
    \item How do artist support networks shape practitioners' experiences and processes? 
\end{enumerate}
\vspace{-0.5em}

We conducted a thematic analysis on collected weekly diary entries and semi-structured exit interviews, situating our results in existing CST theory, paying special attention to emotional well-being and artist-centric benefits \cite{beyondproductivity}. We found that participants often deferred to ArtKrit's suggestions for matters of composition and value, but remained firm in their own color choices even when they did not match the reference image. A quantitative analysis of usage logs revealed that ArtKrit usage decreased over time. As they familiarized themselves with the tool, participants transitioned from early exploratory epistemic actions, which took time, to confidently choosing features that matched with their existing workflows. Creating artist support networks in our evaluation resulted in feelings of creative safety; participants celebrated and were inspired by not how accurately their peers recreated the master studies, but rather their unique personal adaptations. 

Finally, we reflect on methodological implications from this three week, group-based evaluation, where researchers also participated as artists. We argue that CST evaluations can---and should---be structured as opportunities to commission participants to meaningfully engage in their artistic practices. 
Specifically, by intentionally setting a study's normative ground \cite{power-cst}, researchers can co-create a space with participants where they feel artistically and emotionally inspired and supported. \changes{By viewing their role as to set up the infrastructure for participants to meaningfully engage in their practices, researchers concede power to participants to shape the course of the study: } a step towards negotiating and lessening the power differential between researchers and ``human subjects.''  Rather than ``extracting knowledge'' from artists, we hope our study can serve as one example in a broad ``toolbox'' \cite{remy-evaluating-csts} of how CST evaluations can be generative for not just researchers, but also participants---participants who may otherwise be frustrated that their unique, lived experiences extend beyond a narrow frame of knowledge prescribed by research norms \cite{art-not-research}. 

\textit{Note: All the reference artwork in this paper is either public domain,
used with artist permission, or under fair use copyright policy.}

%% file: 02_methods.tex
\begin{table*}[t]
\centering
\renewcommand{\arraystretch}{1.2}
\begin{tabular}{llccll}
\toprule
\textbf{Researcher} & \textbf{Participant} & \textbf{Years of Experience} & \textbf{Age} & \textbf{Digital Tools Used} & \textbf{Artist Identity (Self-Described)} \\
\midrule
\multirow{3}{*}{Researcher 1}
& 1A & 15 & 27 & Photoshop, Procreate & Artist \\
& 1B & 15 & 31 & Autodesk Sketchbook, Pixelorama & Hobbyist \\
& 1C & 10 & 22 & Procreate & Hobbyist \\
\midrule
\multirow{4}{*}{Researcher 2}
& 2A & 8 & 20 & Clip Studio Paint, Photoshop & Professional \\
& 2B & 15 & 22 & Clip Studio Paint, MS Paint, Photoshop & Hobbyist \\
& 2C & 6 & 19 & Procreate & Hobbyist/Professional \\
& 2D & 6 & 18 & Krita, Procreate & Professional \\
\midrule
\multirow{2}{*}{Researcher 3}
& 3A & 7 & 19 & Clip Studio Paint & Hobbyist/Aspiring Professional \\
& 3B & 4 & 20 & Procreate & Hobbyist \\
\bottomrule
\end{tabular}
\vspace{2pt} 
\caption{Participant demographics and self-described artist identities.}
\Description{A table titled "Participant demographics and self-described artist identities" with six columns: Researcher, Participant, Years of Experience, Age, Digital Tools Used, and Artist Identity (Self-Described). The table lists nine participants across three research groups.

Group under Researcher 1: Participant 1A (27 years old, 15 years experience, uses Photoshop and Procreate, identifies as Artist). Participant 1B (31 years old, 15 years experience, uses Autodesk Sketchbook and Pixelorama, identifies as Hobbyist). Participant 1C (22 years old, 10 years experience, uses Procreate, identifies as Hobbyist).

Group under Researcher 2: Participant 2A (20 years old, 8 years experience, uses Clip Studio Paint and Photoshop, identifies as Professional). Participant 2B (22 years old, 15 years experience, uses Clip Studio Paint, MS Paint, and Photoshop, identifies as Hobbyist). Participant 2C (19 years old, 6 years experience, uses Procreate, identifies as Hobbyist/Professional). Participant 2D (18 years old, 6 years experience, uses Krita and Procreate, identifies as Professional).

Group under Researcher 3: Participant 3A (19 years old, 7 years experience, uses Clip Studio Paint, identifies as Hobbyist/Aspiring Professional). Participant 3B (20 years old, 4 years experience, uses Procreate, identifies as Hobbyist).}
\label{tab:participants}
\end{table*}
\vspace{-0.5em}

\section{Methodology}
\label{sec:method}

To answer our research questions, we conducted a longitudinal, mixed-methods user study. Nine participants were grouped into three communities of practice, each including a member of the research team, and used ArtKrit to replicate one master study each week. Participants corresponded with each other and the research team on Discord. In total, the study resulted in 36 artifacts and over 50 hours of ArtKrit usage data. 

\subsection{ArtKrit Improvements}
To support \changes{participants installing ArtKrit on their own devices} during our longitudinal study, we implemented targeted system improvements. First, we reduced latency for adaptive composition line generation, especially on lower-memory machines, by transitioning the adaptive grid feature (the only feature that uses AI models) from a locally hosted model to an online pipeline using Replicate. This change allowed participants to use ArtKrit's adaptive lines without requiring specialized hardware for local runs, which previously required sufficient GPU support and higher memory availability. Second, we integrated incremental Krita saves and usage logging, recording interaction events such as when participants requested feedback for composition, value, or color, along with the full Krita document files saved at corresponding timestamps. These logs provided quantitative traces of tool use over time that we utilized to analyze our participants' drawing processes. 

We also made minor quality-of-life improvements to ArtKrit. Toward more control, users can now regenerate composition lines by manually moving, adding, or subtracting detected key points, without needing to first prompt DINO and SAM. The composition tab additionally enables users to pop out and zoom their reference image. 
Finally, in the color section of the tool, users can choose the background color (previously fixed as white) of the color separation feature, which also can be popped out and zoomed.




\subsection{Participant recruitment}

We recruited nine artists (four women, one man, three non-binary, and one who preferred not to disclose gender) through personal networks, with each researcher largely recruiting participants in their own group. Most of our participants came from snowball sampling where a participant would ask their friend(s) to join the study. All participants were experienced digital artists, though varied in their choice of tools and self-perceived artist identity. Table \ref{tab:participants} shows participant demographics.

We split participants into small communities of practice to maintain a shared sense of cultural identity and context \cite{griffith}. Group one consisted of three CST researchers who were also digital artists; Researcher 1 also shared this identity. Group two consisted of students at an art college---two were roommates, and one of the roommates was friends with the other two---that Researcher 2 had no prior relationship with, but recruited through mutual friends. Group three consisted of an art major at a non-arts focused college and their friend; Researcher 3 was their peer. Group three originally also included an art-major participant who was a friend of Researcher 3's, but withdrew before completing drawings. 

Following Jacobs et al. \cite{para-jacobs}, we compensated participants at rates comparable to commissions; specifically, each participant was paid \$250 USD for three artworks. 

\subsection{Procedure}

\begin{figure*}[htbp]
    \centering
    \includegraphics[width=1\textwidth]{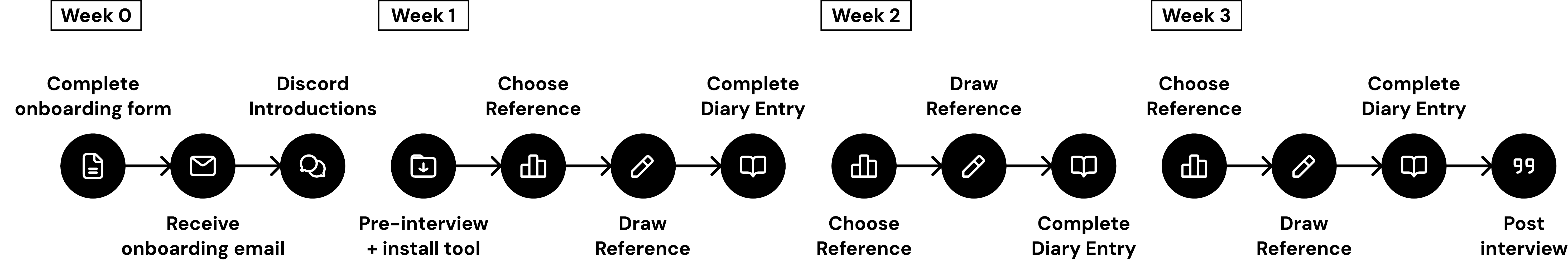}
    \caption{Overview of our evaluation timeline.}
    \Description{A flowchart titled "Overview of our evaluation timeline" showing a 4-week study schedule.
Week 0: Tasks include "Complete onboarding form," "Receive onboarding email," and "Discord Introductions"
Week 1: Tasks include "Pre-interview + install tool", "Choose Reference", “Draw Reference”, and “Complete Diary Entry”
Week 2: Tasks include “Choose Reference", “Draw Reference”, and “Complete Diary Entry”
Week 3: Tasks include “Choose Reference", “Draw Reference”, “Complete Diary Entry”, and “Post Interview”
Arrows connect the tasks sequentially from Week 0 through the weekly cycles to the final Week 3 post-interview.}
    \label{fig:timeline}
\end{figure*}

Our study design is informed by Long et al. \cite{taolongitudinal}, who found that user perceptions of AI tools suffer from novelty effects and undergo a familiarization phase. 
Figure \ref{fig:timeline} shows our evaluation timeline; we conducted the IRB-approved study November to December 2025. While we planned for participants to complete three master studies using ArtKrit over the course of three weeks, in practice the study ran 4-5 weeks due to onboarding and some participants completing Week 3 late as it ran into winter holidays.

First, during virtual onboarding, each researcher helped their group members install ArtKrit on their own computers. The researcher then conducted a 45-minute pre-study interview exploring participants' current practices around completing master studies, sharing art, and giving feedback. 

During each week of the study, each group voted on which reference image to draw, voluntarily shared works-in-progress or other reflections on Discord, reacted to each other's images, and then spent approximately ten minutes completing a private diary entry once their image was done. The diary entry asked about participants' open-ended experiences with ArtKrit and their Discord group. Each facilitating researcher also actively participated by sharing their own artworks and conversing with their group throughout the study, though researchers did not complete diary studies, interviews, or submit usage logs.

Finally, at the end of Week 3, each researcher conducted individual exit interviews with their group members. To see if participants' digital drawing practices, relationships with ArtKrit, or ways of interpreting master studies changed, the exit interviews repeated many questions from the pre-interview, and also included participant-specific elaborations of their behaviors over the past three weeks. Exit interviews also lasted approximately 45 minutes.





\subsection{Data Analysis}
Data collected and analyzed include usage logs, weekly diary entries, Discord messages, and pre- and post-interviews. 
To qualitatively analyze the interview transcripts and diary entries, we conducted a deductive thematic analysis \cite{braun2006using}. We chose a deductive, as opposed to inductive, thematic analysis to explore our pre-existing research questions and to find evidence supporting existing CST theory, such as artist support networks \cite{chung-ast}, dialectical activities \cite{zhang-dialectical-chi2024}, and a lens of power \cite{power-cst}. First, the authors familiarized themselves with the data, and the last author proposed preliminary themes. The first author then coded all participant interviews and developed an initial codebook. While doing so, new themes emerged and were proposed. Through open discussions and weekly meetings, the research team reviewed the interviews and refined the themes.

For quantitative data, we analyzed system usage logs collected during ArtKrit's three-week deployment,  including JSON interaction logs and their corresponding system-logged canvas images. We first cleaned user-uploaded data by removing records outside the three-week study period, and then grouped the data per study group. 
Next, we selected all user-initiated actions and categorized them as composition-related, value-related, or color-related. Finally, we conducted exploratory data analysis in Python to examine trends and usage patterns, including total action counts, the frequency and percentage of each action category, and changes in action usage across weeks and study groups. We additionally analyzed the appearance and disappearance of unique actions over time to understand how participants’ interactions temporally evolved.

\subsection{Positionality}
The first, third, and last authors \changes{(Catherine, Asya, Jingyi)} participated in the study. Originally, we decided to participate mainly to try to reduce attrition through establishing greater rapport with participants, but we found that our experiences in the study also changed our relationships to ArtKrit. While we acknowledge that our actions and reactions cannot be entirely disentangled from those of participants, to retain a narrative that highlights the contributions of participants, we do not intermix our data or impressions with theirs. Instead, we offer first-person accounts and reflections of our experiences below, in line with auto-biographical and self-reflexive methods in HCI \cite{desjardins-autobio}.


\textbf{First author (Researcher 3):} I used Procreate to draw and I uploaded my works in progress to Krita and ArtKrit whenever I was looking for feedback or support. Like many of the other participants, I found the rule of thirds grid the most helpful and used it frequently to create the initial structure of my piece. Previously, I believed that I would’ve preferred to use adaptive composition lines and composition feedback compared to pre-set grids, so I was surprised by how helpful rule of thirds was for me in terms of correct composition.  Before using ArtKrit in our evaluation, I would’ve said that composition, value, and color were equally important. After doing this study I think that composition and value are more significant to the piece and color can be changed according to artist preference. I found myself choosing colors that I liked rather than striving for color similarity with the reference. I also found myself misusing the tool by using the color separation feature for value blocking to show me where lighter and darker areas were. Finally, I would echo something that 1C mentioned, which is that ArtKrit slowed down the initial steps in my workflow, to make future steps easier and less overwhelming. 

\textbf{Third author (Researcher 2):} In the beginning of the study, I did all of my work in Krita. However, towards the end of the first week, I switched to drawing in Procreate and uploading my works in progress onto Krita, similar to my fellow authors. This approach streamlined my drawing process while also getting the necessary support from ArtKrit. I found the composition functions to be the most helpful. I would first generate an adaptive grid and then move the lines around into a position I found the most helpful. This process gave me a good starting point for scaffolding, while also giving me control to create a grid best suited for my needs. I usually don’t focus on values when drawing, but all of my group members emphasized its importance for any piece. Thus, I paid more attention to this aspect and realized that my values tend to be inaccurate. This pushed me to use the value functions more, such as the different filters and kernels. Through these interactions, I discovered a new way to look at drawings and a novel way to use ArtKrit in order to improve my artistic skills. I felt similarly to 2C and 2A who emphasized how the study and its social aspects inspired them to explore different stylistic techniques and ways of using ArtKrit. 

\textbf{Last author (Researcher 1):} Like the first author, I stayed true to my workflow, drawing in Procreate/acrylic and horizontally moving to/from ArtKrit for scaffolds and feedback. I interpreted the master study of Week 1 as fan art of Jacinthe from Pok\'emon: Legends ZA as I often do this in master studies to still feel like I'm retaining creative agency; I didn't consciously know at the time it would set a norm! 
I was inspired by my group's reflections of their creative misuse to push ArtKrit to its limits to see how it would handle traditional art for Week 2. It was tricky to control for ambient lighting while photographing the painting; at first, ArtKrit told me my painting was too red (I had a red underpainting), but in retrospect I think I overcorrected due to the warm lighting of the photo. 
Even though I've used ArtKrit for a long time even before the study, I found my mental framing of it change as a result of my participation. Like what 1C said in their exit interview, ArtKrit was most useful in showing \textit{spatial relationships} of values and color. I realized I now look at and evaluate drawings wholistically, with all aspects of composition, value, and color combined, even if I focus on separate parts during drawing.

As the senior author whose group mates were all more junior HCI researchers, I was always conscious of my position of power and tried not to overstep any bounds (in the exit interviews, group members reported they didn't feel uneven power dynamics). Mainly, though, I was happy to have an excuse to do my own creative practice, and to be supported by artists who felt more like a community of friends than research participants.

%% file: 03_findings.tex
\section{Findings}
\label{sec:results}

Below, we report the results of our three research questions; we supplement our thematic analysis narrative with quantitative usage log analysis. Each subsection's findings are situated in existing CST theory and ends in design implications for other CST researchers and developers. 

\subsection{RQ1: How did participants use ArtKrit in practice?}

\begin{figure*}[!h]
    \centering
    \includegraphics[width=1\linewidth]{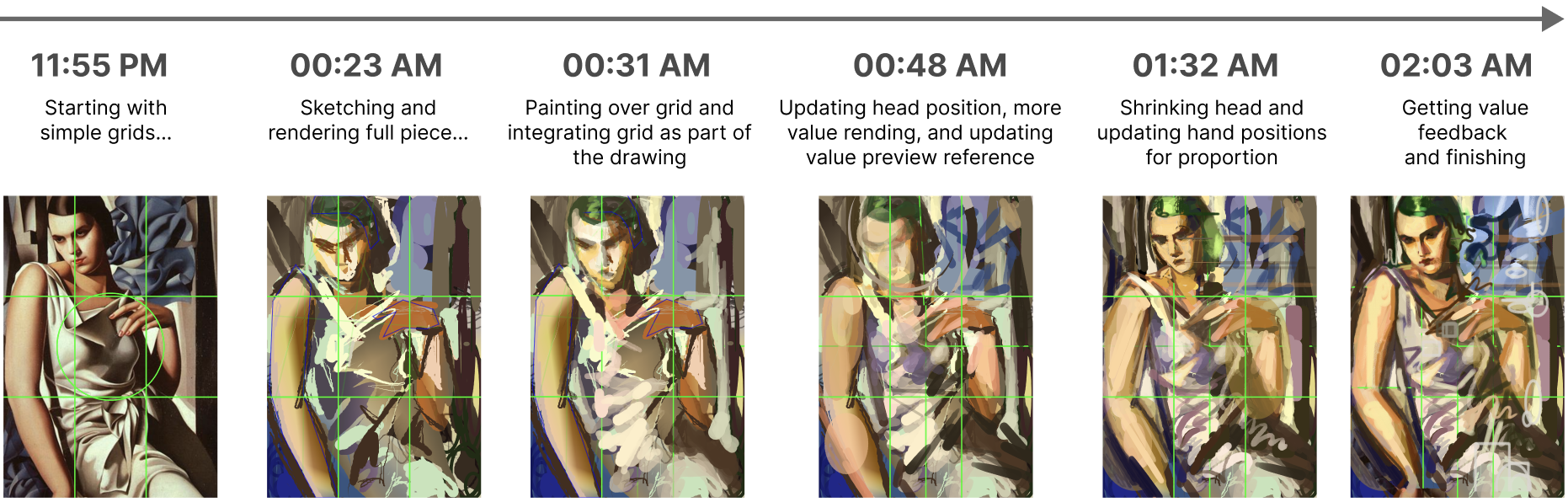}
    \caption{Participant 1A’s three-hour Week 1 drawing session in ArtKrit, showing progression from initial grid setup to final rendering. Key actions include sketching, grid integration, updating previews, refining value, and adjusting proportions.}
    \Description{A timeline log of a digital drawing session, from an initial grid setup to a finished, shaded figure. Key steps shown are sketching, integrating the grid into the drawing, refining light and shadow, and adjusting proportions.}
    \label{fig:artkritUsage1AExample}
\end{figure*}

\begin{figure*}[htbp]
    \centering
    \includegraphics[width=1\textwidth]{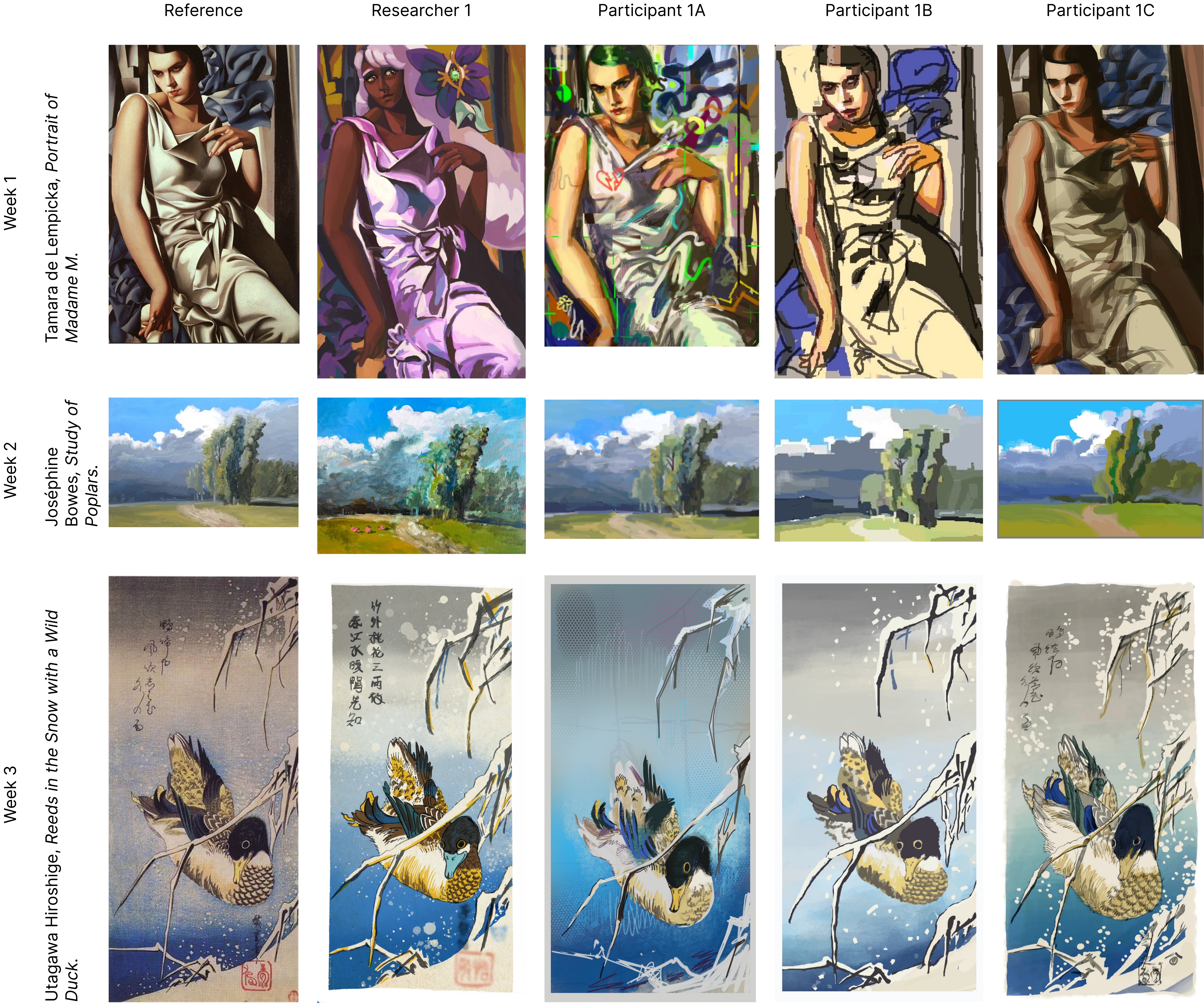}
    \caption{Drawings produced by the participants of Group 1. \textit{Week 1 reference: Portrait of Madame M., Tamara de Lempicka. Week 2 reference: Study of Poplars, Josephine Bowes. Week 3 reference: Reeds in the Snow with a Wild Duck, Utagawa Hiroshige.}}
    \Description{Figure showing a grid of artworks including references and participant drawings across three weeks. Columns are labeled Reference, Researcher 1, Participant 1A, Participant 1B, and Participant 1C. Row 1 (Week 1) presents a portrait reference (Portrait of Madame M. by Tamara de Lempicka) followed by stylized reinterpretations of a seated woman in a white dress, varying in color palette, abstraction, and line quality. Row 2 (Week 2) shows a landscape reference (Study of Poplars by Josephine Bowes) alongside painterly landscape studies of trees, sky, and fields with differing levels of simplification and brushwork. Row 3 (Week 3) displays a Japanese woodblock print reference (Reeds in the Snow with a Wild Duck by Utagawa Hiroshige) followed by participant renditions of a duck among snowy reeds, maintaining the overall composition while varying color saturation, texture, and stylistic detail.}
    \label{fig:group1final}
\end{figure*}

\begin{figure*}[htbp]
    \centering
    \includegraphics[width=1\textwidth]{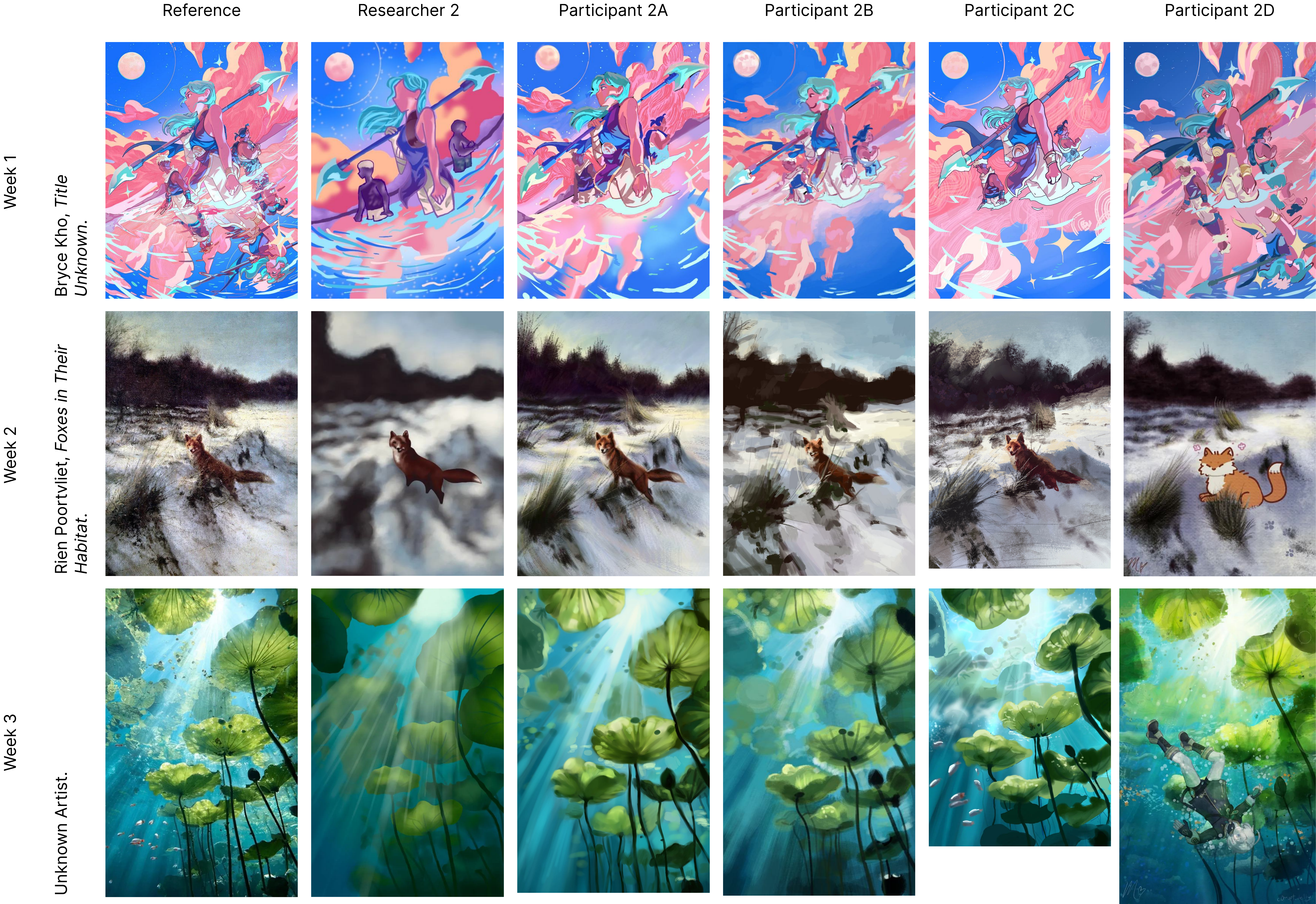}
    \caption{Drawings produced by the participants of Group 2. \textit{Week 1 reference: Title Unknown, Bryce Kho. Week 2 reference: Foxes in Their Habitat, Rien Poortvilet. Week 3 reference, Title Unknown, Artist Unknown.}}
    \Description{Figure showing a grid of artworks comparing references and participant drawings from Group 2 across three weeks. Columns are labeled Reference, Researcher 2, Participant 2A, Participant 2B, Participant 2C, and Participant 2D. Row 1 (Week 1) presents a colorful fantasy illustration reference by Bryce Kho (title unknown), featuring a stylized character and a vibrant sky with pink clouds and blue tones, followed by participant reinterpretations that preserve the dynamic composition while varying character detail, abstraction, and color intensity. Row 2 (Week 2) shows a reference photograph or painting of a fox in a snowy, grassy landscape (Foxes in Their Habitat by Rien Poortvliet), alongside participant drawings that range from painterly and blurred to more illustrative or cartoon-like representations of the fox and environment. Row 3 (Week 3) displays an underwater scene reference (title and artist unknown) with green aquatic plants and rays of light filtering through water, followed by participant interpretations that maintain the vertical plant forms and light beams while differing in texture, saturation, and inclusion of additional elements.}
    \label{fig:group2final}
\end{figure*}

\begin{figure*}[htbp]
    \centering
    \includegraphics[width=1\textwidth]{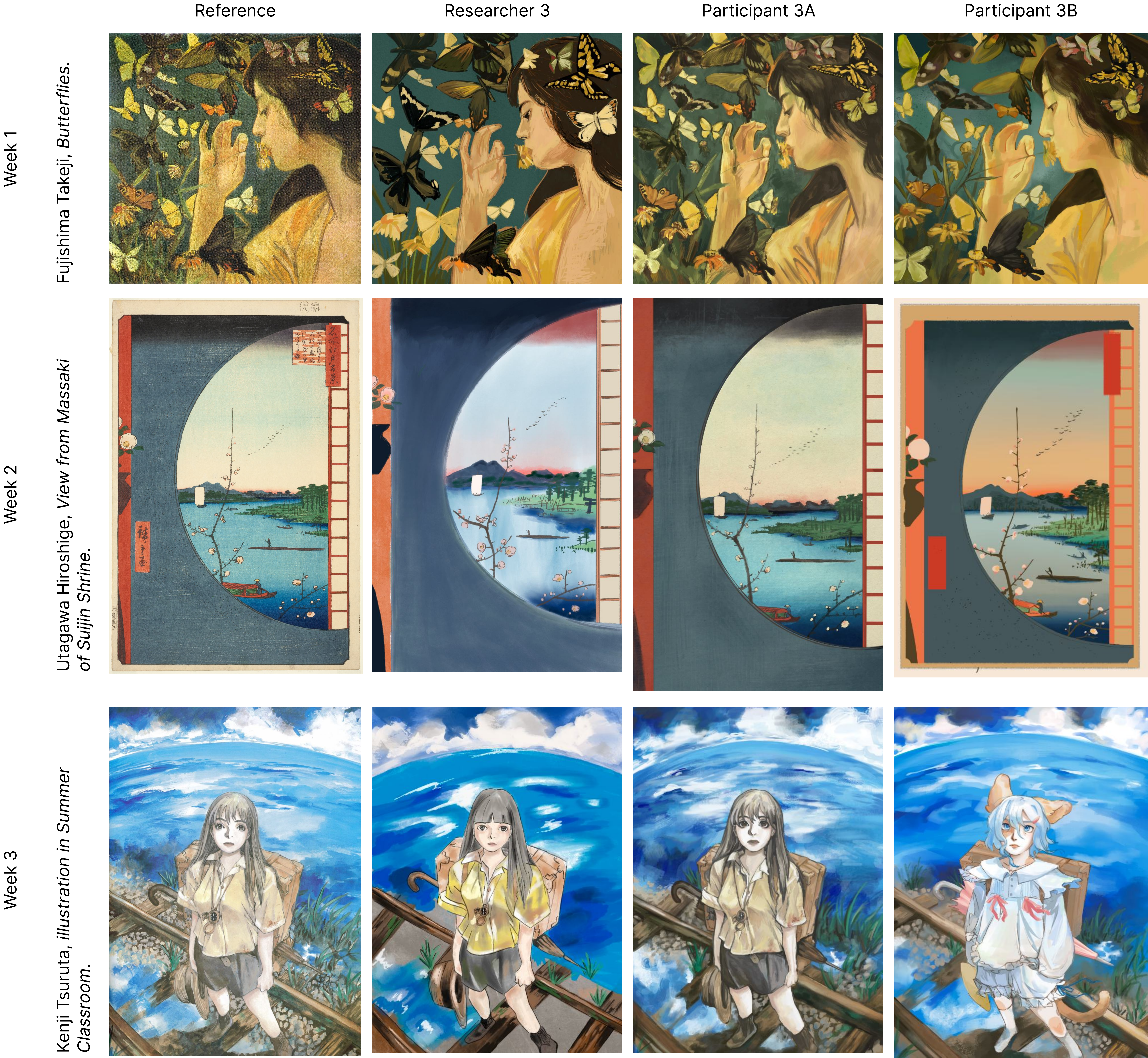}
    \caption{Drawings produced by the participants of Group 3. \textit{Week 1 reference: Butterflies, Fujishima Takeji. Week 2 reference: View from Massaki of Suijin Shrine, Utagawa Hiroshige. Week 3 reference: illustration in Summer Classroom, Kenji Tsuruta.}}
    \Description{Figure showing a grid of artworks comparing references and participant drawings from Group 3 across three weeks. Columns are labeled Reference, Researcher 3, Participant 3A, and Participant 3B. Row 1 (Week 1) presents the reference painting 'Butterflies' by Fujishima Takeji, featuring a soft, impressionistic style with pale colors, followed by participant reinterpretations that explore the composition with varying levels of detail, color palette, and painting style. Row 2 (Week 2) shows the reference 'View from Massaki of Suijin Shrine' by Utagawa Hiroshige, depicting a landscape with water and architecture, alongside participant drawings that reinterpret the traditional scene with different line qualities, color schemes, and digital rendering techniques. Row 3 (Week 3) displays an illustration reference from 'Summer Classroom' by Kenji Tsuruta, featuring a detailed, modern illustrative style, followed by participant interpretations that capture the original's essence while experimenting with stylistic abstraction, color choices, and compositional focus.}
    \label{fig:group3final}
\end{figure*}

On average, participants spent two to three, and up to six, hours completing their weekly master studies. Their engagement with ArtKrit was less than drawing time (Table \ref{tab:weekly_engagement_corrected}). 
Typically, participant workflows started by using grid lines to get roughly sketch out the reference image's composition. Then, using (or misusing \cite{creative-misuse}) various features in color and value, participants would move on to value block or color block. 
Afterwards, in no particular order, participants looked at spatial groupings of color/value from the feedback panel, “vibe checked” their piece, and made continuous adjustments until the piece felt complete. Figure \ref{fig:artkritUsage1AExample} shows an example of Participant 1A’s three-hour drawing session in Week 1. 
Figure \ref{fig:group1final} shows final results from Group 1, Figure \ref{fig:group2final} shows Group 2, and Figure \ref{fig:group3final} Group 3.

\subsubsection{\textbf{Productive Friction}}
Kreminski et al., in Reflective Creators, call for CSTs intentionally designed to elicit reflection and to support ``thoughtful creative practices and practitioners''~\cite{kreminski-reflective-creators}. A key mechanism through which reflective creators prompt reflection is design friction, which interrupts otherwise automatic creative action and invites users to reconsider their choices. We observed multiple instances of reflection emerging from friction introduced into artistic actions, decisions, and interactions with the tool.

One way friction was evoked was through the choices the tool left open to users, rather than the tool making explicit changes. As 1A explained, using ArtKrit required active self-assessment: “you have to actively reflect and be like, am I being supported? Because it's kind of like a thing that's happening [on the sidebar] as opposed to an explicit mark that has changed because you're using ArtKrit.” This ambiguity prompted users to reflect on when to follow the tool’s suggestions and when to assert their own agency.

We observed cases in which participants chose to override ArtKrit’s feedback in favor of their own preferences or creative intentions. For example, after receiving color feedback, both 1C and 3B decided not to adopt the tool’s suggestions. 1C noted, “When I was using ArtKrit and getting feedback on the colors, I wasn't usually taking it, because I know what colors I like, even though it doesn't match the reference image.” Similarly, 3B described ignoring a prompt because replicating the figure exactly wasn’t their intention for Week 3: “the feedback it gave me was like, you don't have enough yellow. And I was like, yeah, I know that. Because like my finger isn't wearing a yellow shirt, so obviously.” In these moments of friction, artists made a reflective decision about their judgment taking priority over the tool’s judgment. This tension is especially salient in the context of master studies, where the goal is often to replicate an existing piece. However, we observed artists acknowledging and ignoring ArtKrit's feedback when replication did not contribute to their learning or artistic goals, such as when they were instead focused on personal expression.

At the same time, we also saw cases where participants prioritized ArtKrit’s feedback. 3B reflected, “I was a little bit more dismissive of some of the things that ArtKrit was telling me that I was doing wrong, but I would say that values and composition were definitely still in effect.” When participants acted on this feedback, they reflected on whether doing so supported their artistic learning. 1B, for example, adjusted the placement of their duck after receiving composition feedback, but noted that “it didn't require [them] to actually understand visually where the duck was; [they] just took what the tool was telling [them] at face value,” which made the experience “feel like tracing.” Across these examples, we see a recurring tension between questions of support (“am I being supported by the tool?”) and growth (“am I actually improving as an artist?”). 

We observed a pattern across these instances: we saw user agency take priority in color, and tool suggestions take priority in value and composition. Participants such as 3A, 3B, and others emphasized maintaining the core essence, or most significant elements, of the reference image as their goal when completing master studies. This leads us to ask: were participants more willing to heed ArtKrit’s feedback on composition or value because they believed these elements were most essential to maintaining the essence of the original work?

\subsubsection{\textbf{Dialectical Activities}}
\label{sec: dialectical}
Zhang defines dialectical activities as activities whose values are rooted in the intrinsic experience of engaging in the activity itself, rather than in external outcomes ~\cite{zhang-dialectical-chi2024}, citing examples such as art-making and other creative pursuits. Zhang further calls for ecosystems that simultaneously produce desired goods and services while also promoting people’s sustained engagement in dialectical activities. In our study, we observed instances where ArtKrit appeared to shape and intensify dialectical engagement within the art-making process. Rather than supporting instrumental goals (e.g., producing an accurate image), ArtKrit introduced meaningful moments of reflection, emotional responses, and perceptual considerations. 

ArtKrit’s features encouraged practices rooted in learning visual fundamentals, which improved participants’ ability to artistically see. One of ArtKrit's most frequently used features across all groups was toggling the rule-of-thirds grid. Several participants noted that simply pressing a button to see grid lines, rather than drawing them manually, prompted them to engage in practices they might otherwise skip or overlook. Similar ``single button press'' actions included viewing a black-and-white version of the canvas (for values) instead of manually applying a grayscale filter, and hovering over regions in the color separation tool rather than manually color-blocking. 2A explained, “I don't usually compare values. It's kind of difficult when you're drawing from real life, or it's just kind of tedious, but ArtKrit made it pretty easy to see all these different aspects of your reference that you wouldn't normally think about.” 2C added, “I wouldn’t really think much about it, but looking at it now, I’m kind of noticing the style of the artist and the colors, how the colors relate to each other, and how they really put time into the composition of the piece… before, I would just mindlessly draw without thinking about any of those things.” 1A described this process as “eating your vegetables”: “Having [ArtKrit] do [things] for me in a way that fed me my vegetables, I see why this is a nourishing thing to do.” ArtKrit’s support in improving seeing highlights one type of dialectical value that is intrinsic to artistic creation.

Similarly, ArtKrit can provide emotional and cognitive support in the art creation process by making complex images feel more approachable and less overwhelming. Alleviating these barriers is another way that using ArtKrit can support dialectical practices. 3A reflected on Week 1's image, explaining that ArtKrit helped them “understand the fundamentals a lot better, rather than being overwhelmed” by the many butterflies in the image. 1A described a similar experience for Week 3’s duck study, highlighting how the intersection of the adaptive grid lines encompassed a specific region of feathers that they could focus on rendering, rather than feeling overwhelmed by the whole duck. 2A emphasized that ArtKrit gave them the tools to “dissect” color and value, enabling them to engage more mindfully with individual elements of the piece. 

\subsubsection{\textbf{Reference Complexity Shapes Tool Use}}
Reference image complexity influenced ArtKrit use. 
For more complex paintings, such as the Group 3's Week 1 \textit{Butterflies} study, ArtKrit was integrated directly into the process of placing elements onto the canvas. In contrast, for simpler pieces, such as the Group 1's Week 2 \textit{Study of Poplars}, ArtKrit functioned more as a post-hoc “sanity check.” For example, 1A used ArtKrit to test their color accuracy only after that portion of the piece was already complete, framing it more as a game to check their color matching abilities. Participants exercised agency in deciding both the timing and extent of feedback, engaging with ArtKrit at different stages of their workflow. Overall, simpler pieces led artists to prioritize the cohesion of the whole composition, whereas more complex pieces encouraged attention to individual sections. This was evident in Group 2’s complex Week 1 image, which 2B described as a “paint by numbers” that they were mindlessly copying because the colors were so complex.


\subsubsection{\textbf{Surfaced Limitations}}

Through our three week deployment of ArtKrit, we also identified several limitations. Because ArtKrit is designed for replicating reference images, it is less effective for original work or intentional deviations from a reference. For example, 3B had to dismiss some of ArtKrit’s suggestions due to intentional differences from the reference in Week 3. As a workaround, Researcher 1, who also added original aspects to their studies, edited the reference image to reflect desired changes. Several participants also noted performance issues dissuading them from using features, particularly the slow generation of adaptive composition grid lines, which took between thirty seconds and one minute on average on the Replicate servers. Additionally, 2D did not use the color feature because they found it complicated and felt that the value feature was sufficient. Participants offered suggestions for improvement as well: 1B expressed interest in a more localized value/color selection feature, which would allow comparisons of smaller regions rather than the entire canvas. As part of our commitment to iteratively improving the tool with participants, we have updated ArtKrit to add this functionality.

\subsubsection{\textbf{Design Implication: Integrated Simplicity Bests Novel Algorithms}}
Throughout the study, we observed that sometimes, the most technically simple features (i.e., rule of thirds grid) had the greatest influence on artistic practices, especially when compared to novel AI-assisted algorithms (i.e., adaptive grid lines). 
Even though all participants were capable of manually drawing a rule of thirds grid in about a minute, the ability to press a button to instantly generate one was enough of a savings in manual labor that participants could instead focus on observation and interpretation, rather than spending time creating scaffolds. Just because a feature is known, easy to implement, and not a technical advancement does not mean it will not be impactful to creative practices. 

\subsection{RQ2: How did relationships with ArtKrit change over time?}


\subsubsection{\textbf{From Exploration to Targeted Use}}
As participants used ArtKrit over the course of three weeks, their relationship with the tool evolved. During the first week, participants engaged in more exploratory use, experimenting with different parts of ArtKrit to understand how it could support their existing practices, foregrounding epistemic over pragmatic actions~\cite{tricaud-beaudouinlafon-epistemic-creative-behaviour}. Early on, 1A described ``\textit{going around the different tabs}'' and observing the cognitive impact of various features. By the end of the study, participants had settled into using the features that best supported their individual workflows.

\input{figure/results_activeUsageTime}


An analysis of usage logs also reflects this shift from exploratory to targeted use.
To understand users’ active usage time with ArtKrit over three weeks, we computed active hours and days for each user and group from system logs. We defined active time as consecutive actions within a 30-minute window (e.g., users might be working on drawing between system interactions) and an active day as any calendar day on which a participant performed at least one action. Table~\ref{tab:weekly_engagement_corrected} shows the elapsed active hours and days each participant spent working with ArtKrit over the three weeks. Across all nine participants, they spent an average of 1.89 active hours over 1.94 calendar days per week.
Across weeks, the average time spent shows a clear declining trend in two groups: Group 1 (from 2.49 hours to 1.33 hours, then to 1.10 hours) and Group 3 (from 3.45 hours to 2.59 hours, then to 1.61 hours).  The decline in usage time of ArtKrit can potentially be attributed to exploratory actions in using a new tool: participants engaged more deeply with ArtKrit in the first week to understand how the tool could fit into their existing workflows, after developing mental models of how it could be used, they utilized only features they knew to be helpful for them ~\cite{tricaud-beaudouinlafon-epistemic-creative-behaviour}. This idea is expanded upon in \ref{sss:familiar}.

\begin{figure}[t!]
    \centering
    \includegraphics[width=\linewidth]{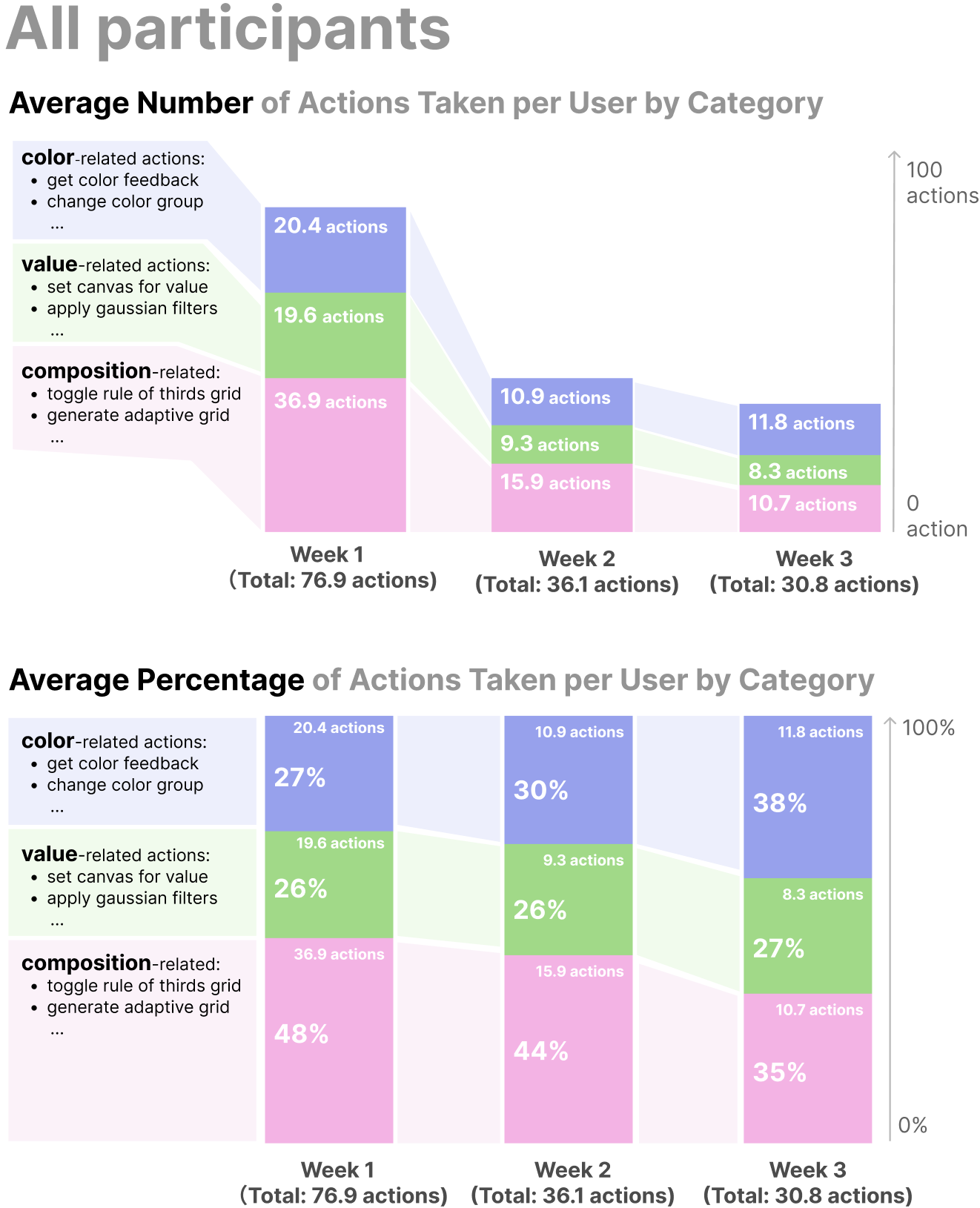}
    \caption{Average number and percentage of user actions by category (color, value, composition) over three weeks.}
    \Description{Figure shows two stacked bar charts comparing user actions across three categories over three weeks. The top chart displays the average number of actions, while the bottom chart shows the same data as percentages. Both charts use three colors: blue for color-related actions (such as get color feedback and change color group), green for value-related actions (such as set canvas for value and apply gaussian filters), and pink for composition-related actions (such as toggle rule of thirds grid and generate adaptive grid). In Week~1, users completed 76.9 total actions: 20.4 color actions (27\%), 19.6 value actions (26\%), and 36.9 composition actions (48\%). Week~2 showed 36.1 total actions: 10.9 color actions (30\%), 9.3 value actions (26\%), and 15.9 composition actions (44\%). Week~3 had 30.8 total actions: 11.8 color actions (38\%), 8.3 value actions (27\%), and 10.7 composition actions (35\%). The data shows a declining trend in total actions over the three weeks, with composition-related actions decreasing most significantly while color-related actions maintained a relatively higher proportion by Week~3.}
    \label{fig:results_actionCountsTotal}
\end{figure}

Besides decreased active usage time, participants’ interactions with ArtKrit also declined over the course of the study in terms of both the total number of actions taken and the breadth of unique actions used. As shown in the top chart in Figure~\ref{fig:results_actionCountsTotal}, the average total number of actions dropped from 76.9 in Week~1 to 36.1 in Week~2 and further to 30.8 in Week~3, representing a 53.1\% decrease from Week~1 to Week~2 and an overall 59.9\% decrease from Week~1 to Week~3. Similar to trends observed in earlier sections, this reduction likely reflects a shift away from exploratory behavior toward more targeted use of ArtKrit features. This overall decline was consistent across groups, as shown in all three left charts in Figure~\ref{fig:result_actionCountByGroup}, though the trajectories differed: Groups 2 and 3 monotonically decreased across weeks, while Group 1 had more activity in week 3 compared to week 2, likely due to the increased complexity of the piece. 


\changes{In addition to taking fewer actions overall, participants also engaged with a smaller set of unique action types over time. After the first week, users performed fewer distinct actions on average when using ArtKrit---on average, about four action types less across all groups (a 19\% decrease).} 
Over time, participants developed an understanding of which features supported specific aspects of their workflow, and learned to draw on different parts of ArtKrit as needed, such that “when they cognitively needed that support, they knew which tab of ArtKrit to go to.” (1A). Similarly, mental models of ArtKrit’s potentials and limitations emerged as the study progressed, suggesting sustained and effective use across its feedback dimensions rather than increasing reliance on a single aspect. For example, when using adaptive composition lines, participant 1B initially thought that text inputting a single object would cause lines to appear only through that object, but later realized that prompting two objects was necessary to generate lines between them. Taken together, these patterns suggest that participants gradually developed more streamlined and personalized workflows with ArtKrit, relying on a smaller but stable subset of features as they gained experience across tasks.

\vspace{-1em}

\subsubsection{\textbf{Convergence Toward Balanced Use Across Color, Value, and Composition Actions}}
Beyond this shift from exploratory to targeted use, participants' engagement across ArtKrit's three feedback dimensions—color, value, and composition—also became more balanced, with an initial imbalance that gradually equalized over time. As shown in the summative bottom chart in Figure~\ref{fig:results_actionCountsTotal}, participants’ interactions were dominated by composition-related actions (48\%) in Week 1, with color-related (27\%) and value-related actions (26\%) used substantially less. Over the following weeks, however, these differences narrowed. Composition-related actions decreased, color-related actions increased, and value-related actions remained relatively stable, before converging toward a more even split. Initial imbalances may stem from participants’ prior drawing habits, stylistic preferences, early exploration of ArtKrit's novel composition algorithm, or the specific reference images used in a given week. As participants continued using ArtKrit, they increasingly converged toward more balanced usage across three aspects rather than reliance on a single dominant category.

\begin{figure*}[!h]
    \centering
    \includegraphics[width=.9\linewidth]{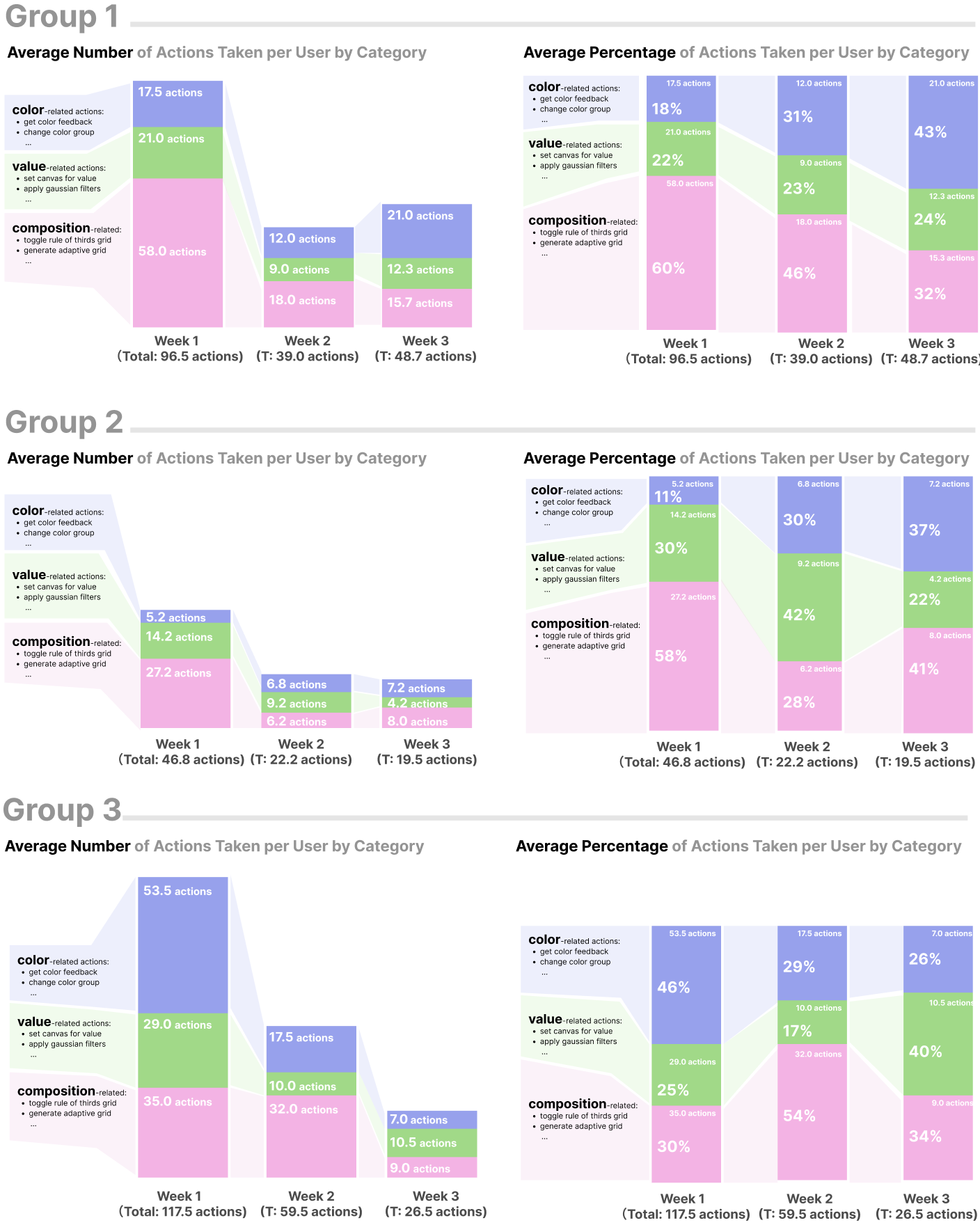}
    \caption{Average number and percentage of user actions by category (color, value, composition) over three weeks by each study group.}
    \Description{This figure displays user actions across three study groups over three weeks, with each group showing two stacked bar charts: average number of actions (left) and average percentage of actions (right). All charts use blue for color-related actions, green for value-related actions, and pink for composition-related actions.  

Group~1 shows declining total actions from Week~1 (95.8 actions) to Week~2 (39.0 actions) to Week~3 (48.7 actions). Week~1 comprised 17.5 color actions (18\%), 21.0 value actions (22\%), and 58.0 composition actions (60\%). Week~2 had 12.0 color actions (31\%), 9.0 value actions (23\%), and 18.0 composition actions (46\%). Week~3 showed 21.0 color actions (43\%), 12.3 value actions (24\%), and 15.3 composition actions (32\%).  

Group~2 also declined from Week~1 (46.8 actions) to Week~2 (22.2 actions) to Week~3 (19.5 actions). Week~1 had 5.2 color actions (11\%), 14.2 value actions (30\%), and 27.2 composition actions (58\%). Week~2 showed 6.8 color actions (30\%), 9.2 value actions (42\%), and 6.2 composition actions (28\%). Week~3 had 7.2 color actions (37\%), 4.2 value actions (22\%), and 8.0 composition actions (41\%).  

Group~3 decreased from Week~1 (117.5 actions) to Week~2 (59.5 actions) to Week~3 (26.5 actions). Week~1 comprised 53.5 color actions (46\%), 19.0 value actions (25\%), and 35.0 composition actions (30\%). Week~2 had 17.5 color actions (29\%), 10.0 value actions (17\%), and 32.0 composition actions (54\%). Week~3 showed 7.0 color actions (26\%), 10.5 value actions (40\%), and 9.0 composition actions (34\%).  

All groups show significant decreases in total actions over time, with varying distributions across the three action categories.}
    \label{fig:result_actionCountByGroup}
\end{figure*}

All three study groups showed similar equalization patterns, as shown in the three right charts in Figure~\ref{fig:result_actionCountByGroup}. Before equalizing, Group 1 showed a strong imbalance toward composition in Week 1; Group 2 also initially favored composition though used ArtKrit the least of all groups; in contrast, Group 3 favored color actions Week 1 and had the biggest decline (77\%) in overall ArtKrit usage from the start to end of the study. 

\subsubsection{\textbf{Creative Misuse and Evolving Artistic Practices}}

We also observed instances of what might be characterized as creative misuse, where participants employed ArtKrit in ways beyond its original design intent ~\cite{creative-misuse}. In general, participants became more comfortable misusing the tool after becoming familiar with it, following their initial phase of exploration. For example, 1C stated that ArtKrit's most helpful feature was the neon contours that highlighted similar regions of value or color. While these were originally designed for feedback between the canvas and reference image, 1C repurposed these visually salient elements to guide their composition. They noted that the shapes of the neon blobs helped them see sharper regions of color, and used it when color blocking their drawings. 
Another example of creative misuse involved 1A incorporating ArtKrit’s neon green grid lines directly into the artwork (Figure \ref{fig:artkritUsage1AExample}) as their goal was to ``respect the medium of ArtKrit,'' describing them “as if there were green laser lights in the room with [the figure],” or treating ArtKrit as an intuition training tool for checking their color matching skills in Week 2. Cases of creative misuse may be attributed both to participants' own agency while interacting with ArtKrit and to the community as participants were inspired by their peers' practices. 

Through participating in the evaluation, participants reported changes in their artistic practices, preferences, and skills that they intended to carry forward beyond the study. For example, 2D discovered new brushes they wanted to continue using: “[The master studies] made me use a lot more brushes, like they put me out of my comfort zone. I actually found a bunch of new options [that] I'm probably going to be using a lot in the future.” Similarly, 1A described feeling more prepared to pursue physical plein-air landscape paintings after completing Week 2's \textit{Study of Poplars}.

\subsubsection{\textbf{Design Implication: Reaffirming the Familiarization Phase}}
\label{sss:familiar}

Our results suggest that participants’ use of tools evolves as their familiarity with them increases. Over time, they begin to integrate the tool into their existing workflows and practices. For example, 3B, a participant in an earlier evaluation of ArtKrit, noted that in the first study they experimented more consistently with ArtKrit’s features as the study took place in a lab setting and using ArtKrit was what they were there to do. In contrast, in the current study, 3B was more comfortable using their usual setup and was more selective about the features they found most useful. We offer evidence in support of Long et al.'s theory that users undergo familiarization phases with novel tools \cite{taolongitudinal}. We believe that as epistemic, exploratory actions account for Week 1's high usage counts, CST evaluations that argue for a tool's usability should consider actions after a familiarization phase to avoid novelty effects.  

\subsection{RQ3: How did artist support networks shape participants' experiences?}
\label{sec:asn}

\subsubsection{\textbf{Community for Emotional Well-Being}}

Using ArtKrit within an artist support network \cite{chung-ast} contributed to participants’ emotional well-being through positive interactions and encouragement. Consistent with theories of social support and communities of practice, these interactions functioned as expressive and affirmational support, reinforcing participants’ sense of belonging and legitimacy within the group ~\cite{wenger-communities-of-practice}. Feedback and discussion within groups were almost entirely positive, with participants commenting on the uniqueness of each master study and individual expression rather than accuracy or similarity to the original. Some representative Discord comments reacting to finished pieces include, “I like that soft line brush you used for the clouds” (2A reacting to 2D’s Week 1 painting) and “This is adorable, I love the outfit” (2C reacting to 3B’s Week 3 painting), in addition to the messages in Figure \ref{fig:teaser}. Completing studies within a group was also grounding for participants. As Wenger describes, communities of practice “provide homes for identities,”~\cite{wenger-communities-of-practice} allowing participants to situate their struggles and progress within a shared practice. 2C shared that they enjoyed having a space to talk with others about the pieces rather than getting “in their own head;” when they were struggling and heard 2A mention similar difficulties, it made them feel that their struggles were seen. 3B added that being part of such a group felt almost equalizing: when viewing professional artists’ work, it can be difficult for them to imagine how the piece was made, but seeing peers successfully complete the same studies was both reassuring and motivating. In viewing participants' encouragement as a reception activity in a creativity support ecosystem \cite{cse-shm}, we highlight how peer feedback and affirmation extend beyond evaluation, functioning instead as social infrastructure that supports emotional resilience and creative persistence. 

\subsubsection{\textbf{Community Driven Motivation and Learning}}

Erickson and Kellogg highlight the importance of systems that allow users to “see” one another, make inferences about others’ activities, and imitate one another, arguing that such social translucence enables environments to “support the same sort of social innovation and diversity that can be observed in physically based cultures” ~\cite{erickson-kellogg-social-translucence}. Through our longitudinal group evaluation, participants described becoming more motivated, more willing to invest effort, and more likely to explore and integrate new possibilities into their own workflows through observing and responding to others’ practices. For instance, 1C went back to and spent an additional three hours on their duck drawing in Week 3, citing they did not want it to ``look bad'' compared to their peers' drawings, which may explain the increased average actions and active days shown in Table \ref{tab:weekly_engagement_corrected} and Group 1's Week 3 bar in Figure \ref{fig:result_actionCountByGroup}.

We observed how participants were influenced in both managing- and teaching-type tasks in their artist support network \cite{chung-ast}. For instance, for managing-type tasks, participants noted that seeing the stage others were at during the week motivated them either to get started or to continue working. Participants continually set and reacted to standards of effort and completeness: 2B mentioned that seeing some people “turn in” minimal pieces was reassuring, as it affirmed that even if their own piece was not amazing, it was still acceptable. 1B noted that “seeing what details other people chose to spend time on rendering definitely influenced [their] choices on what to spend time on, or how long to keep going,” effectively setting a benchmark. They also mentioned learning through observing others in teaching-type tasks: “I definitely picked things up from seeing other people do it, and then I think it also, like, gave me an idea of what quality bar to shoot for.” 1A similarly indicated that they wanted to put in effort because they saw others doing so, “not in a pressuring way, but in a co-working sense.” 2C, who initially relied on eyedropping colors, copied less over time and instead tried to “learn” more by observing others’ work.

\subsubsection{\textbf{Setting Normative Ground}}

A group-based evaluation also helped expand what was possible within the study. Many participants pointed Researcher 1's modified Week 1 master study (Figure \ref{fig:teaser}) as a realization of what could be done, which led to increased personalization and content changes in participants’ work in subsequent weeks. Under a lens of power \cite{power-cst}, researcher participation helped set a normative ground by example, rather than researchers explicitly enforcing or suggesting practices. Beyond researcher involvement, participants’ own practices also inspired the practices of others. For instance, 1A’s inclusion of green grid lines in Week 1 inspired 3B to do something similar in week three, and 1B’s use of a trackpad encouraged 1A and 1C to do their Week 2 studies with a trackpad as well. As 1A put it, “you don’t really see the possibilities until you see somebody kind of pushing the possibilities.”


\subsubsection{\textbf{Fun!}}

Lastly, all participants reported that they had fun participating in the study: “I had so much fun! I wanna do more studies! I think a Discord group where people make art together is a very good way to do a study!” (1A); “Working with the tool was really fun. It felt exciting to be part of making something new,” (3A); and “I really had fun with it. Yeah, I really enjoyed the study,” (2D). Beyond completing study procedures, this underscores the often overlooked aspect of evaluations including user-centric value, whether through positive emotions, increased self-confidence, or overall well-being ~\cite{beyondproductivity}. Our findings reveal the potential of CST evaluations to contribute positively to participants' lives through lived experiences, highlighting the value of considering affective and experiential outcomes as part of understanding how creative systems are experienced by users.

\subsubsection{\textbf{Design Implications: A Push for Greater Group Studies}}

These findings highlight the value of social and collaborative structures in supporting creative work. Being part of an artist support network not only encouraged emotional well-being through positive feedback and shared experiences, but also helped establish implicit standards for motivation, effort, and the possibilities of practice. Artists create art, in large part, to react to the norms around them \cite{creative-misuse}, and a group-based study presents an opportunity for any participant to enact a new norm. We believe group-based studies may expand the space of what is artistically possible while providing support and emotional-well being for creative practices, and encourage CST researchers to look into adapting group-based methodologies to evaluate their own tools.

%% file: figure/results_activeUsageTime.tex
\begin{table*}[!h]
\centering
\begin{tabular}{lcccc|cccc}
\hline
     & \multicolumn{4}{c|}{Avg. Active Hours (per user)} & \multicolumn{4}{c}{Avg. Active Days (per user)} \\
        \cline{2-4} \cline{5-7}
      &  3-Weeks Avg. & Week 1  & Week 2  & Week 3  & 3-Week Avg. & Week 1  & Week 2  & Week 3  \\
\hline
1A & 1.64 & 2.91 & 1.78 & 0.22 & 2.91 & 2.00 & 2.00 & 1.00 \\
1B & 1.97 & 2.07 & 1.65 & 2.21 & 2.33 & 2.00 & 2.00 & 3.00 \\
1C & 0.70 & n/a & 0.55 & 0.86 & 2.00 & n/a & 2.00 & 2.00 \\
\textbf{Group 1 Avg \#} & \textbf{1.44} & \textbf{2.49} & \textbf{1.33} & \textbf{1.10} & \textbf{2.00} & \textbf{2.00} & \textbf{2.00} & \textbf{2.00} \\
\hline
2A & 1.73 & 1.43 & 2.73 & 1.02 & 1.67 & 2.00 & 2.00 & 1.00 \\
2B & 1.72 & 0.66 & 2.87 & 1.62 & 2.00 & 2.00 & 2.00 & 2.00 \\
2C & 0.38 & 0.42 & 0.04 & 0.68 & 1.33 & 1.00 & 1.00 & 2.00 \\
2D & 3.73 & 6.10 & 4.17 & 0.93 & 2.00 & 3.00 & 1.00 & 2.00 \\
\textbf{Group 2 Avg \#} & \textbf{1.89} & \textbf{2.15} & \textbf{2.45} & \textbf{1.06} & \textbf{1.75} & \textbf{2.00} & \textbf{1.50} & \textbf{1.75} \\
\hline
3A & 3.87 & 5.15 & 3.30 & 3.16 & 2.33 & 3.00 & 2.00 & 2.00 \\
3B & 1.23 & 1.75 & 1.87 & 0.07 & 2.00 & 2.00 & 2.00 & 2.00 \\
\textbf{Group 3 Avg \#} & \textbf{2.55} & \textbf{3.45} & \textbf{2.59} & \textbf{1.61} & \textbf{2.17} & \textbf{2.50} & \textbf{2.00} & \textbf{2.00} \\
\hline
\textbf{All Participants} & \textbf{1.89} & \textbf{2.56} & \textbf{2.11} & \textbf{1.20} & \textbf{1.94} & \textbf{2.14} & \textbf{1.79} & \textbf{1.90} \\
\hline
\end{tabular}
\vspace{5px}
\caption{Average weekly active hours and days per participant in ArtKrit. Note: Week 1 data logs for Participant 1C are unavailable because they were not uploaded and were accidentally deleted by the participant.}
\Description{Table showing average active hours and active days per user in ArtKrit application across three weeks. Data is organized by three groups (1, 2, and 3) with individual participants labeled A, B, C, or D. The table has two main sections: Average Active Hours per user and Average Active Days per user, each showing a 3-week average and weekly breakdowns for weeks 1, 2, and 3. Group 1 includes participants 1A, 1B, and 1C with a group average of 1.44 active hours. Group 2 includes participants 2A, 2B, 2C, and 2D with a group average of 1.89 active hours. Group 3 includes participants 3A and 3B with a group average of 2.55 active hours. Overall, all participants averaged 1.89 active hours and 1.94 active days across the three-week period. Note indicates Week 1 data for Participant 1C is unavailable because data logs were not uploaded and were cleaned by the participant.}
\label{tab:weekly_engagement_corrected}
\end{table*}

%% file: 04_discussion.tex
\section{Discussion}
\label{sec:discussion}

Here, we reflect and speculate on our evaluation methodology and its implications for future HCI research. While our findings revealed that researchers have the power to set a study's normative ground, some may view this intentional setting with concern as it could ``bias'' participants. We offer a response, and argue that evaluations---not just the research phase of tool design---are still sites for creative practice. We believe CST researchers should use their power, authority, and financial resources to enact ad-hoc artist support networks during evaluations.  

\subsection{Setting normative ground versus biasing participants}
In section \ref{sec: dialectical}, we argued that our longitudinal group evaluation of ArtKrit made space for dialectical activities from which participants derived some intrinsic benefit. Participants' behaviors were in part guided by the normative ground researchers set, such as Researcher 1 turning a study into Pok\'emon fan art to show personal interpretation is accepted---and even welcomed. 

Traditional HCI evaluations, such as controlled in-lab studies, may frame norm setting as a threat to validity or neutrality, with the potential to bias participants. However, we argue that norm setting is essential to CST evaluations---particularly in evaluating CSTs with artistic goals, where participants value the creative process and being expressive within wide walls \cite{resnick-principles}. In a creative practice, norms shape what kinds of learning and expression are possible. Artists create art, in part, to respond to norms \cite{creative-misuse}. Design probes are not value neutral \cite{gaver-cultural-probes}, and, echoing calls of defining what the \textit{design} goals of a CST should be \cite{hewett2005creativity,frich_mapping_2019,remy-evaluating-csts}, HCI researchers should also define their \textit{evaluation} goals. By offering examples that sit outside of participants' pre-conceived notions of acceptable master studies for a research study, researchers ``biased'' participants to expand their notions of the space available for creative opportunity. We hope this resulted in more meaningful creative engagements, such as how participants then commented on and supported each other's unique touches, which bolstered creative safety and led to more creative risk taking. 

Thus, we argue for moving towards thinking of CST evaluations not as data gathering in service of validating a tool's design goals, but rather as \textbf{infrastructure design}. In designing the structure of an evaluation, researchers have the opportunity to guide participants emotionally, epistemically, and practically through meaningful creative engagements. For instance, even though 2A said they would not post any of their master studies on social media since none of them were in their usual fan art style, they still cited developing transferable skills through participating in the study. Though spiritually similar to participatory design methods \cite{bodker2022participatory}, which emphasize co-designing the \textit{tool}, none of our participants aided in the design of ArtKrit. Instead, the \textit{study} was a co-created experience. While researchers initially planned the structure (e.g. small communities of practice that draw the same master study each week), participants helped recruit one another into the community of practice, selected the master studies, and established feedback norms. With just researcher guidance, participants might not have been sufficiently motivated and influenced in their artistic choices, such as how the rest of Group 1 mirrored 1B's track pad drawing during Week 2, rather than Researcher 1's traditional painting. Rather than a shared ownership of tools with participants, designing CST evaluations as infrastructure for artist support networks results in a shared ownership of a creative experience. 

To frame this discussion under a lens of power \cite{power-cst}, we distill two design implications:

\begin{enumerate}
    \item By also participating as artists, researchers may intentionally set normative ground through demonstrating their own artifacts.
    \item Framing CST evaluations as building (1) infrastructure for artist support networks and (2) opportunities for meaningful creative engagement is a way researchers can mitigate the power imbalances they have with participants.
\end{enumerate}

\subsection{Generalizing to other CSTs}
While the goals and affordances of ArtKrit dictated some of our study structure, we believe our methodological lessons learned generalize to other CSTs. A CST evaluation is one of the rare opportunities researchers have to evaluate real artists using their tools---under the Creativity Support Ecosystem \cite{cse-shm} perspective, group-based longitudinal studies mean that researchers have some control over distribution and reception activities, not just creation ones. And the distribution and reception activities that occur in an evaluation are also parts of the creative practice: \textbf{a CST evaluation is not external to the design process; it is one of the sites where creativity is shaped, practiced, and made legible.} The participants researchers recruit and the tasks researchers delegate all are ways to shape the resulting ecosystem and artist support network. It is our hope that CST evaluations consider this wholistic view, as the HCI community works towards creating generative sites of understanding human creative activity. 

Granted, a longitudinal group-based evaluation is just one tool in the toolbox of potential CST evaluations \cite{remy-evaluating-csts}, and evaluations should be picked to match the goals of both researchers and participants. A longitudinal group-based technology probe that aims to uncover emergent patterns of creative use may not be the best fit for proving metrics like increased efficiency or reduced cognitive load or tests of statistical significance. 
\changes{It is true that ArtKrit's focus on disciplined drawing (as opposed to original work) and all groups drawing the same image limits the ecological validity of our artifacts. 
At the same time, these artifacts afforded comparison within and between groups, embodying a design position that draws inspiration from controlled experiments while still accounting for socio-cultural factors. }
When researchers organize participants into groups paying attention to their cultural backgrounds \cite{griffith}, and empower participants to decide which artifacts to create, they may receive, in return, \changes{chances to observe creative and cultural divergences}. As demonstrated in section \ref{sec:asn}, evaluations that enact artist support networks may result in participants pushing themselves to create more polished artifacts. In contrast to Phraselette's evaluation \cite{phraselette-10.1145/3715336.3735832}, which was also situated in existing sites of creative practice (i.e. poetry writing), ArtKrit evaluation's increased Discord activity potentially may have been due to the accountability of not wanting to let other participants, or the researchers, down. 

Finally, if researcher-artists also participate in the study, in addition to setting norms, they too can experience personal well-being benefits. For instance, Researcher 1 learned about Giuseppe Castiglione (also known as Lang Shining), a 17th century Italian Jesuit painter/Chinese missionary who blended East-West styles, when their group was deciding which image to pick for Week 3. Finding out about an ancient painter who drew Qianlong Emperors as if they were Napoleon felt meaningful to their personal artistic practice, even if the group did not decide to do his piece. Every member of the research team who participated felt grateful to have external accountability to draw again: even if the study had not resulted in any interesting insights, at least the research team had an excuse to give time and care to their neglected artistic practices. We hope that CST evaluations can be an opportunity for researchers to use their power to bring together participants in community, orchestrating meaningful opportunities for artistic engagement.




\subsection{Limitations \& Future Work}

Though we argue that CST evaluations can be opportunities for meaningful creative engagement, they are still research studies. A research study has its own normative ground. It fails to capture other aspects of artistic milieu that are important in artist support networks, such as different audiences on social media, while it also enacts new norms, such as how institutional affiliation may cause participants to behave in correct, expected ways. Though the researchers took great care to communicate that participants may choose to use or not use ArtKrit however they want in accordance to their existing practices, more work is to be done about the refusal of not just tools, but also the refusal of norms. Even if researchers intentionally set norms they believe to be positive for participants, how can they create a space for participants refuse to engage with such norms without being socially ostracized by their peers? How can researchers show up for their responsibility to, as Satchell and Dourish say, ``take people seriously'' \cite{beyond-the-user}? How can researchers balance the idiosyncratic preferences, processes, and ethics of individual artists with their desire to remain in community? 

We did not attempt to isolate independent variables, as our evaluation was not a controlled study but a technology probe toward generating a wholistic view of how relationships with technology change temporally and socially. In addition, effects of ArtKrit may be confounded with the social infrastructure surrounding the study. This makes it difficult to isolate impacts of the tool and community-driven influences. More specific causal conclusions may strengthen this and future studies. For instance, when a participant changed their practice, was it more due to fellow participants' behaviors, or more due to their interactions with ArtKrit? How much exactly did researcher participation influence the study outcomes? We believe the value of rigorous measurement metrics in design studies lie not in justifying the usability of a tool, but in disentangling the sources of authority that govern behaviors in researcher-enacted creativity support networks and ecosystems. 

Our study design also had several limitations. Our participant sample was mostly young and entirely North American based. Our small Discord groups were purposefully grouped by shared backgrounds; future work could run small groups with participants from widely different backgrounds to see if emotional support and creative safety replicates, or if participants can influence each other with even more idiosyncratic practices. Though a deployment, our study was not fully an in-situ study (note that we use this term rather than ``in-the-wild'' in response to Ssozi-Mugarura et al.'s critiques of the colonial origins of the phrase \cite{enough-in-the-wild}). Researchers grouped up participants, who all had to draw the same master study, often different than their usual ``style,'' even if the image was chosen via group consensus. While a few participants did post their pieces on social media, a vast majority of them---particularly self-described professionals---chose not to because they do not share their studies in the first place, opting instead to solely post fan art to build their social media presence. A true in-situ study could recruit Krita users from forums with little to no guidance on the master studies used with ArtKrit---if users even choose to draw master studies at all.



%% file: 05_related_work.tex
\section{Related Work}

While this paper detailed one tool in a toolbox of potential CST evaluations \cite{remy-evaluating-csts} in an attempt to shift the norms of how such evaluations are conducted, many other researchers have evaluated---or struggled with evaluating---CSTs.

\subsection{Evaluating Creativity Support Tools}

CSTs span a wide variety of goals, from supporting divergent thinking, to producing novel artifacts, to expanding the possible space of artistic expression~\cite{frich_mapping_2019}. 
CSTs typically can help reduce users’ cognitive and temporal effort when co-creating artifacts, such as a written piece or an illustration~\cite{longTweetorialHooksGenerative2023a}. As a result, evaluations of CSTs have primarily focused on rigorous comparisons, often using a baseline in which participants create artifacts with or without the system, and assessing user experience, task performance, or the quality of the resulting creative outputs, sometimes through expert or peer annotation of those artifacts~\cite{taolongitudinal,10.1145/3746058.3758469}. Common user evaluation approaches of CSTs utilize standardized metrics such as the Creativity Support Index (CSI)~\cite{CSI}, NASA Task Load Index (NASA-TLX)~\cite{nasa-tlx}, and the System Usability Scale (SUS)~\cite{brookeSUSQuickDirty} and rely on quantitative analyses of interaction time and user engagement~\cite{beyondproductivity}. 
These approaches provide methodological rigor and allow relatively straightforward comparisons across tools or conditions, offering insights into usability and performance benefits. \changes{More recently, comparative structured observation \cite{CSO-mackay} offers methodological rigor in qualitative assessment of design concepts rather than usability goals. }
    
Prior work has raised concerns that artifact-centric metrics and ad-hoc, usability- or productivity-focused evaluation frameworks for CSTs fail to capture the complexity, diversity, and subjectivity of creative practice~\cite{remy-evaluating-csts, taolongitudinal}.
Creativity is often messy, exploratory, and deeply personal, making it difficult to evaluate using standardized or outcome-oriented measures such as effectiveness or efficiency~\cite{beyondproductivity}. As a result, existing evaluations may overlook how CSTs support long-term creative development, evolving user behaviors and interactions, and experiences that extend beyond a single user or single session~\cite{beyondproductivity, taolongitudinal, blaynee2016collaborative} shared there is limited work that examines how CSTs are appropriated, adapted, or abandoned over time in real-world creative practice. \changes{To remediate this, work in research through design has recently called for using medium fidelity prototypes to create formative situations \cite{formative-situations} or to create a shared vocabulary of design events to better capture temporality \cite{design-events}.} Our work builds on these critiques and approaches by exploring an alternative evaluation that prioritizes in-situ use over time and emergent creative practices over short-term performance metrics~\cite{remy-evaluating-csts, coughlan2009understanding}.

\subsection{Longitudinal Studies and Community-Based Evaluation Studies}
To complement traditional in-lab, one-off, individual-user studies, evaluations of HCI tools can extend beyond temporal and social constraints, capturing longer-term use and community dynamics.
For the temporal aspect, longitudinal research in HCI emphasizes observing technology use over extended periods to capture learning, appropriation, and evolving practices~\cite{remy-evaluating-csts,nielsen_longitudinal_2021,userstudy_consideredharmful}. In real-world deployments, long-term use can reveal user breakdowns, workarounds, and evolving user needs or usage patterns that are often invisible in short-term evaluations~\cite{10309195,jain2012user,sporka2011chanti}. Additionally, extended in-situ usage can uncover issues such as novelty effects~\cite{fitbit_novelty,novelty_microsoft}, learning curves~\cite{moran_everyday_2002, berker2005domestication}, customization and personalization~\cite{Muller_Neureiter_Verdezoto_Krischkowsky_AlZubaidi-Polli_Tscheligi_2016,10.1145/2157689.2157804}, as well as misuse, nonuse, and appropriation~\cite{DesigningForAppropriationAlanDix, 10.1145/3613904.3642242, creative-misuse}. The researchers behind tools such as Para \cite{para-jacobs} and Phraselette \cite{phraselette-10.1145/3715336.3735832} have conducted extended user studies that prioritize artistic agency and real-world creative practices; we are inspired by their methods, though they do not explore group-based evaluations. 

For the social aspect, community-based studies further highlight how social interaction and learning, peer feedback, and shared norms shape how tools are used, interpreted, and valued~\cite{10.1145/2998181.2998195}. When a new tool is introduced, community dynamics—shaped by both past practices and emerging behaviors—evolve over time, revealing how adoption, adaptation, and collaborative use unfold~\cite{Kraishan2025TheAAA}. Community-based system evaluation enables more nuanced analyses of interactions beyond the one-user-one-tool dynamic, closer to real-world scenarios~\cite{Aragon2009ATOA,10.1145/3746058.3758469}. Our evaluation of ArtKrit addresses this gap by modeling a community-based, longitudinal evaluation of a creativity support tool situated within an artist support network \cite{chung-ast} and creativity support ecosystem \cite{cse-shm}. 

%% file: 06_conclusion.tex
\section{Conclusion}
To understand how artists work with digital drawing creativity support tools over time and within artist communities, we conducted a three-week longitudinal evaluation of ArtKrit with nine participants organized into three communities of practice. Our study revealed that participants initially explored the tool broadly through epistemic actions but gradually narrowed in on features that supported their individual workflows. 
Participants took ArtKrit's feedback on matters of composition and value but stuck with their choices for color. Qualitative findings highlighted dynamics within artist support networks, where participants engaged in positive feedback and discussion, reporting that these interactions made them feel seen, reassured, motivated, and accountable. Participants had fun drawing in community with each other: our study contributes a methodology for future CST evaluations, demonstrating they can be structured as meaningful, generative opportunities for artistic practice.